\documentclass[twocolumn, times, tighten]{aastex61}
\usepackage{amsmath,amstext}
\usepackage[percent]{overpic}
\usepackage[figure,figure*]{hypcap}
\usepackage{newtxmath} %use times font for math

\shorttitle{Observing galaxy mergers at the epoch of reionization}
\shortauthors{Chaikin et al.}

\defcitealias{Bouwens2015UVFIELDS}{B15}

\begin{document}

\title{Observing galaxy mergers at the epoch of reionization}
\author[0000-0003-2047-3684]{Evgenii A. Chaikin}
\affiliation{Peter the Great St. Petersburg Polytechnic University, Politechnicheskaya 29, 194021, St. Petersburg, Russia}
\author[0000-0003-1826-5149]{Nadezda V. Tyulneva}
\affiliation{Peter the Great St. Petersburg Polytechnic University, Politechnicheskaya 29, 194021, St. Petersburg, Russia}
\author[0000-0003-0255-1204]{Alexander A. Kaurov}
\affiliation{Institute for Advanced Study, Einstein Drive, Princeton, NJ 08540, USA}

\correspondingauthor{Alexander A. Kaurov}
\email{kaurov@ias.edu}

\begin{abstract}

  The galaxies with photometric redshifts observed in a close angular proximity might be either projection coincidences, strongly lensed images of the same galaxy, or separate galaxies that are in a stage of merging. We search for the groups of galaxies in the Hubble Ultra Deep Field (HUDF09) in $z\sim7$ and $z\sim8$ drop-out samples. We find no close pairs among 50 galaxies in the $z\sim7$ sample, while in the $z\sim8$ sample we find that 6 out of 22 galaxies have a companion within $\sim1\arcsec$  (3 pairs).  
  Adopting a numerical simulation and performing forward modeling we show that even though mergers are unlikely to have such a high fraction, the projection coincidences and the strong lensing are even less likely mechanisms to account for all of three pairs. Alternatively, there is a possibility of the contamination in the drop-out catalog from lower redshifts, which potentially can account for all of the groups.
  Finally, we make projection on the sensitivity to mergers of the James Webb Space Telescope, and discuss the possible applications of the high-redshift merging galaxies for decreasing cosmic variance effect on the luminosity function and for improving the accuracy of photometric redshifts in general.
\end{abstract}


\section{Introduction}
The next decade will be rich for new space observatories sensitive to the infrared, such as the James Webb Space Telescope \citep{Gardner2006TheTelescope} and the Wide Field Infrared Survey Telescope \citep{Spergel2015Wide-FieldReport}, that would allow us to greatly extend the observational horizon well into the epoch of reionization, and the ground-based instruments like the Giant Magellan Telescope \citep{Johns2012GiantOverview} would provide greater angular resolution in infrared. The main statistics that will be immediately extracted from the deep galaxy surveys is the luminosity function and the average star formation rate \citep[e.g.][hereafter B15]{Bouwens2015UVFIELDS}.
In this paper we study another piece of information that can be potentially contained in the existing and upcoming imaging  data -- the statistics of galaxy mergers at the epoch of cosmic reionization.

The fraction of mergers is well studied at the redshifts below $\sim 4$ \citep[i.e.][]{Lotz2011THE1.5,Man2016,Conselice2008,Lopez-Sanjuan2009}, and was compared with simulations \citep[i.e. recent comparison with the Illustris simulation ][]{Snyder2017}. See \citet{Rodriguez-Puebla2017TheGyrs} for a combined review of available data on merger observations. Also, the fraction of clumpy galaxies was studied up to redshifts $8$ \citep[see compilation of various studies in][]{Shibuya2016MORPHOLOGIESGALAXIES}. The large number of both spectroscopic and photometric galaxies at low redshifts allows one to make magnitude cuts and study major/minor mergers separately. At the redshifts of reionization, on contrary, there are much less observed galaxies, especially spectroscopically confirmed. Nevertheless, in this paper we attempt to see what is observed at $z\sim7-8$ and how consistent is it with the numerical simulations.

The galaxies at redshifts beyond $6$ are expected to be compact and their detailed morphology can not be presently resolved \citep[e.g.][]{Liu2017Dark-agesGalaxies}. However, individual blobs (that appear to be gravitationally bound) can be observed. There is already a handful of Ly$\alpha$ emitters (LAEs) observed at high redshifts that exhibit either multiple clumps, merging, or have another galaxy nearby:
``Himiko'' object has three clumps that may correspond to a triple merger at $z\approx7$ \citep{ouchi_discovery_2009,ouchi_intensely_2013}; 
``CR7'' (COSMOS Redshift 7) at $z=6.6$ that exhibits three distinct clumps with significantly different photometric SEDs \citep{sobral_evidence_2015}; 
A1689-zD1 is a potential dusty merger at $z\approx7.5$ \citep{watson_dusty_2015, knudsen_merger_2016}; the lensed object at $z=6.3$ that has two distinct clumps and three images \citep{rydberg_multiply-imaged_2016}.

However, Ly$\alpha$ line is detected only for a fraction of objects. In the absence of spectroscopic redshifts, we have only photometric estimates that have much lower precision and accuracy especially at high redshifts. Therefore, the projection uncertainty fundamentally limits the search of grouped of objects. Nevertheless, we show that there is still some signal that can be extracted. In this paper we focus on the Hubble Ultra Deep Field 2009 \citep[HUDF09, ][]{Beckwith2006TheField} and adopt the Hubble eXtreme Deep Field legacy data (XDF, \citet{Illingworth2013TheEver}) and consider drop-out samples of Lyman-break Galaxies at $z\sim7$ and $z\sim8$ from \citetalias{Bouwens2015UVFIELDS}. We show that even without spectroscopic redshifts there is a substantial statistical signal in the data (\S\ref{sec:obs}), and using one group as an example we argue that, indeed, it can be identified as one system (\S \ref{sec:example}). 

The theoretical considerations \citep{Lacey1993MergerFormation} also predict a significant fraction of dark matter halos and, therefore, galaxies to be in some stage of merging. In order to compare theoretical predictions with observations, we perform forward modeling using numerical simulations of cosmological size boxes with galaxy formation (\S\ref{subsec:numsim}). We mimic the HST's Wide Field Camera 3 (WFC3), including noise and point spread functions, and generate mock observations with a depth similar to the XDF (\S\ref{subsec:mimobs}). The mock images are then processed with common pipelines and source catalogs are created (\S\ref{subsec:proc}). In \S\ref{sec:results} we summarize the results of our modeling and compare it with other possible mechanisms that can create observed grouped galaxies. Finally, in \S\ref{sec:discussion}, we discuss the possible application of the grouped galaxies.

\section{Observations of mergers}
\label{sec:obs}
\begin{figure*}[!t]
\begin{center}

\textit{
F775W\;\;\;\;\;\;\;\;\;\;\;\;\;\;\;\;\;\;\;\;\;\;\;\;\;\;
F850LP\;\;\;\;\;\;\;\;\;\;\;\;\;\;\;\;\;\;\;\;\;\;\;\;\;\;
F105W\;\;\;\;\;\;\;\;\;\;\;\;\;\;\;\;\;\;\;\;\;\;\;\;\;\;
F140W\;\;\;\;\;\;\;\;\;\;\;\;\;\;\;\;\;\;\;\;\;\;\;\;\;\;
F160W}

\includegraphics[height=120pt]{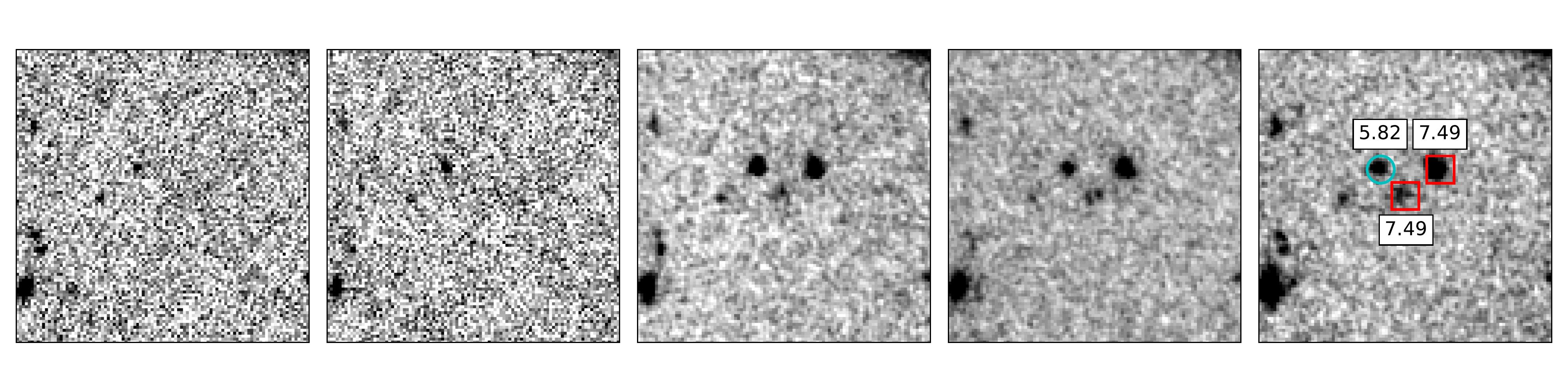}
\includegraphics[height=120pt]{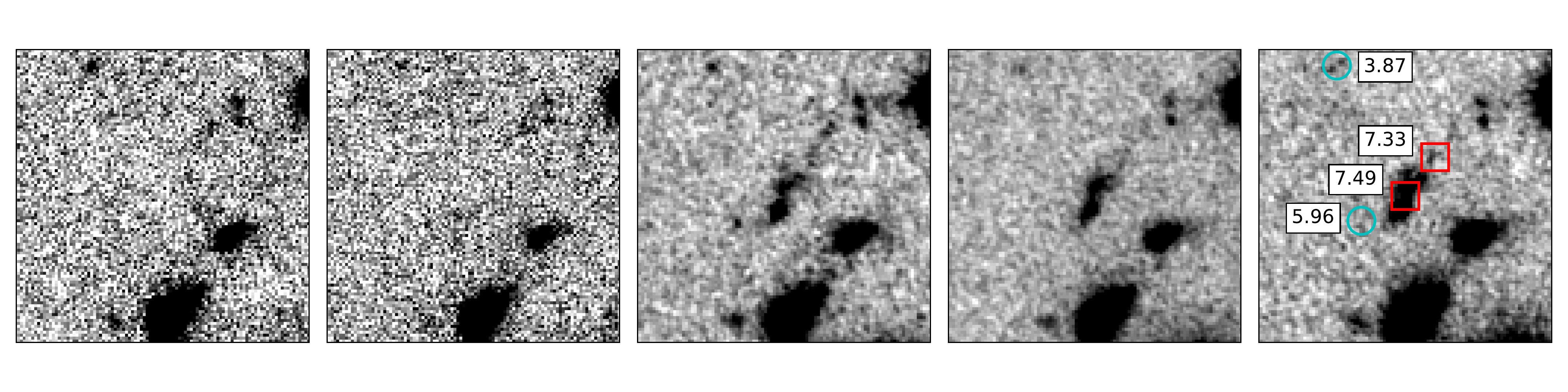}
\includegraphics[height=120pt]{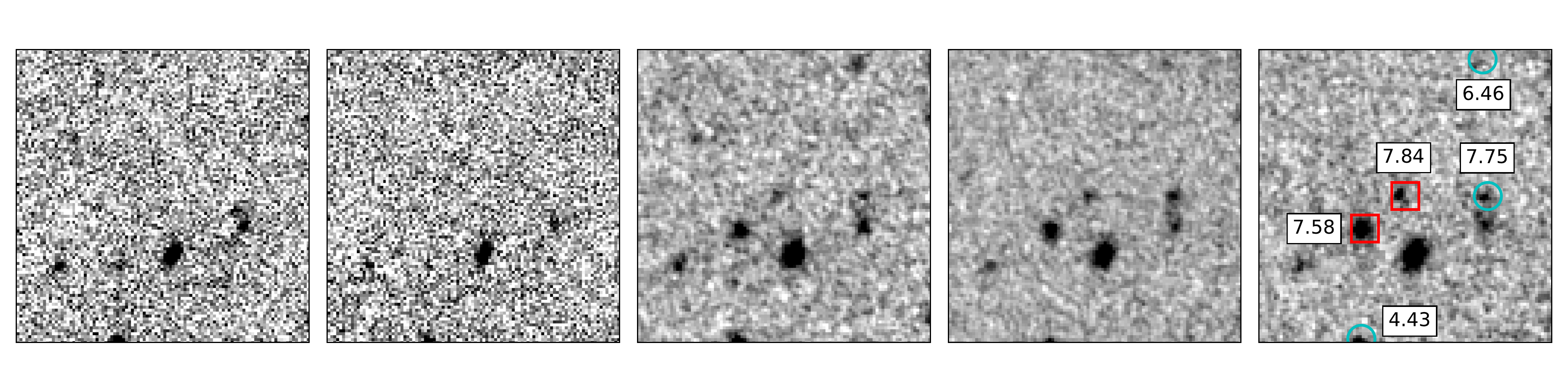}

\end{center}

\caption{\label{fig:cutout}
Cutouts of galaxies from $z\sim8$ (red squares) drop-out sample with separation less than $3\arcsec$. Only galaxies with $z>4$ are labeled with cyan circles and corresponding photometric redshift. All panels are 6 arcseconds across, and the colormap in all panels spans from $-1\sigma$ to $+2\sigma$. The galaxy in the last cutout with photometric redshift $z=7.75$ belongs to $z\sim7$ drop-out sample. Coordinates of groups from up to bottom (03:32:44.7; -27$^{\circ}$ 46$'$ $44.8\arcsec$), (03:32:40; -27$^{\circ}$ 47$'$ $17.5\arcsec$), (03:32:43; -27$^{\circ}$ 46$'$ $27.3\arcsec$). 
}
\end{figure*}

\begin{figure*}
\begin{minipage}{0.35\textwidth} %
\hspace*{0.8cm} \includegraphics[width=0.75\columnwidth]{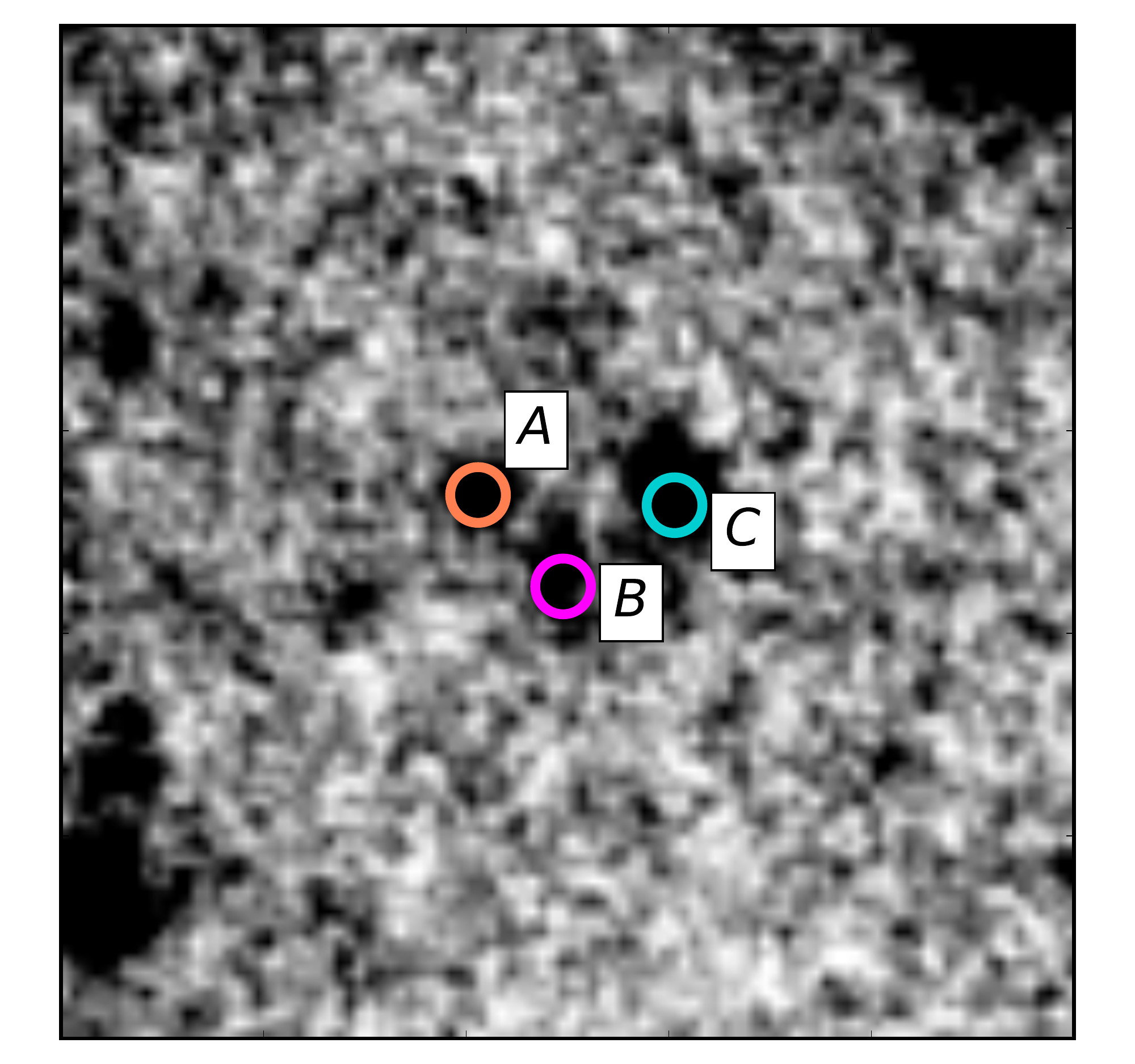} \\
\includegraphics[width=0.9\columnwidth]{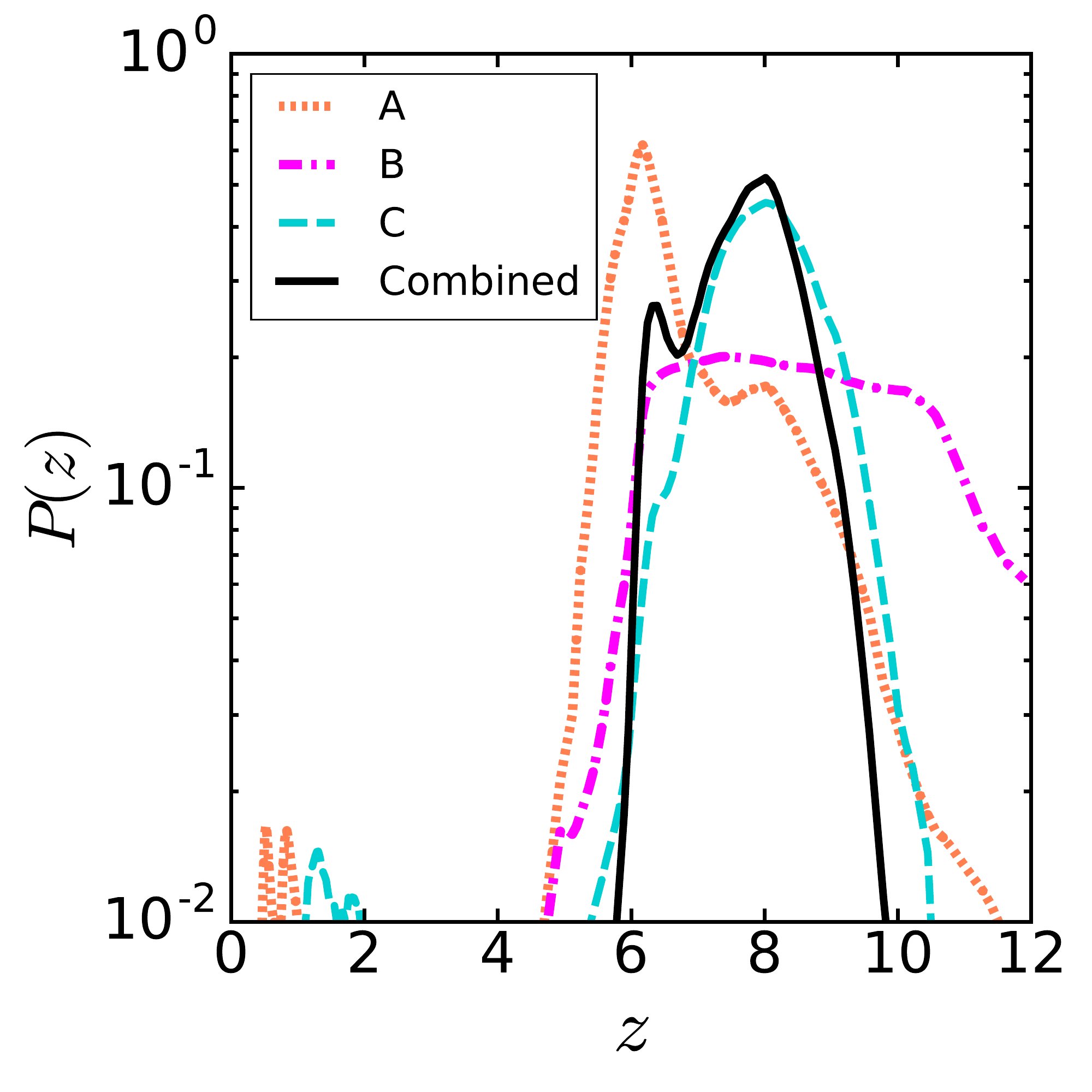}
\end{minipage} %
\begin{minipage}{.65\textwidth} %
\includegraphics[width = 1.01\columnwidth]{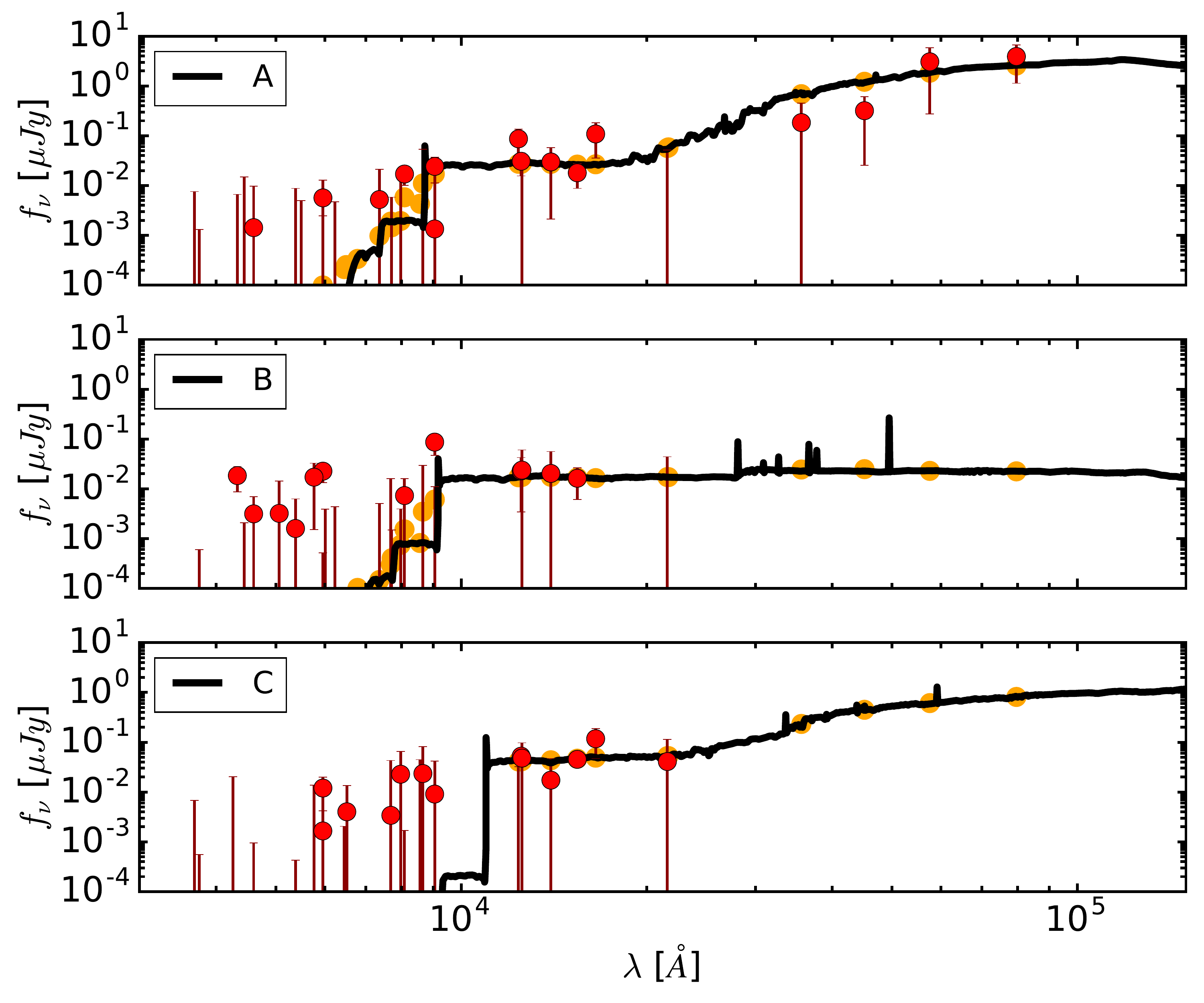}
\end{minipage}
\caption{\label{fig:sed_pz}\textit{Top left panel:} Cutout (3:32:47.9; $-27^{\circ}46\arcmin53\arcsec$) of a group from the science image combined from F125W, F140W, and F160W filters \citep{Momcheva2016THEGALAXIES} with size of 6 arcseconds across (colormap spans from -3$\sigma$ to 2$\sigma$). This system corresponds to the first row in Figure \ref{fig:cutout}. \textit{Right panel:} Spectral energy distributions for each object. Solid lines are the best fit evaluated by EAZY code \citep{Brammer2008EAZY:Code}, yellow points are theoretically predicted flux values and red points are the actual photometric data taken from 3D HST survey \citep{Momcheva2016THEGALAXIES}. \textit{Bottom left panel:} Corresponding redshift probability distribution $P(z)$ for each of three objects (color-coded) and combined probability (solid line).}
\end{figure*}

\begin{figure}
\includegraphics[width=0.99\columnwidth]{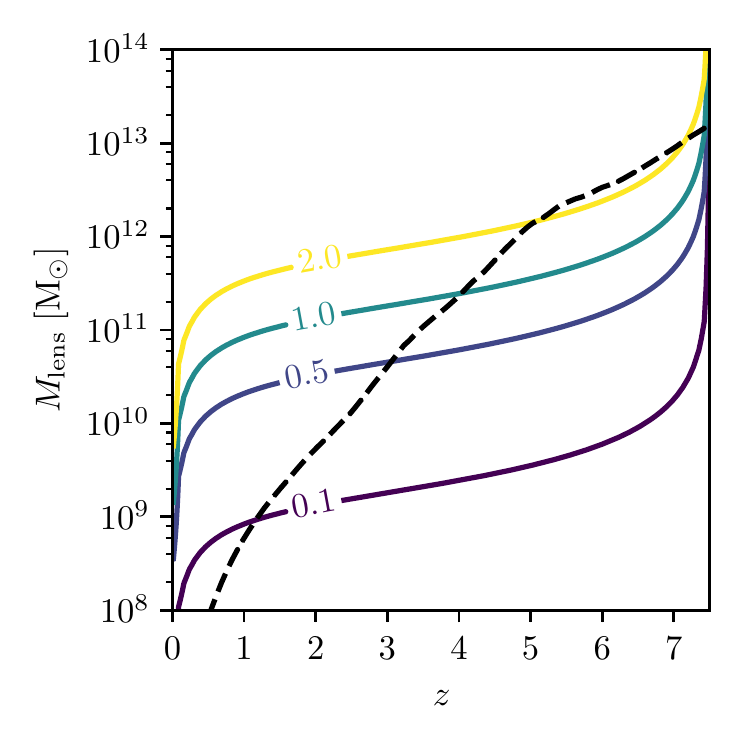} 
\caption{\label{fig:lens}The maximum separation in arcseconds for two images (solid lines) of an object at $z\approx7.5$ for a lens at given redshift and with given mass. Detection limit of unresolved lenses for HST WFC3 in $F160W$ filter (dashed line) in assumption of an elliptical galaxy SED \citep{Polletta2007SpectralSurvey} and $M/L_B \sim 5$, following the procedure used in \citet{Tsai2015THEWISE}.  }
\end{figure}

In this study we consider the HUDF09 and the XDF project data release, since it is likely to be the deepest imaged region until the JWST will start taking measurements. Also, this field is well imaged by ground based telescopes and the photometry is conveniently available \citep{Brammer20123D-HST:Telescope}. The Frontiers fields (with images of high-redshift galaxies magnified by foreground clusters) effectively provide even deeper imaging; however, for this study we avoid the complications caused by lensing effects.

We adopt the source catalog used to determine the luminosity function at $4<z<10$ in \citetalias{Bouwens2015UVFIELDS} which is available online\footnote{\url{http://cdsarc.u-strasbg.fr/viz-bin/Cat?J/ApJ/803/34}}. Another catalog with open access -- the HST 3D survey \citep{Skelton20143D-HSTMASSES} -- also covers this field, but it differs by the way sources are detected, as well as the photometry measurements and final photometric redshifts. We do not use the 3D HST redshift catalog since by construction it assumes the upper limit for the photometric redshifts to be 6, i.e. the objects that are determined as $z>6$ in the \citetalias{Bouwens2015UVFIELDS} catalog are $z<6$ in the 3D HST catalog. Also, there is a catalog by \citet{Mclure2011} that covers this region and redshift range. We discuss it in \S\ref{sec:res:contamination}.

The galaxies in \citetalias{Bouwens2015UVFIELDS} catalog have two redshifts. First one is the photometric redshift evaluated with template fitting code EAZY \citep{Brammer2008EAZY:Code} using the photometric fluxes in different filter bands. Second is the redshift bin based on drop-out technique. We adopted $z\sim7$ and $z\sim8$ samples that correspond to the drop-outs from F850LP and F105W HST filters \citep{Bouwens2011ULTRAVIOLETOBSERVATIONS}. By its nature, the photometric and drop-out redshifts are much less precise than spectroscopic redshifts and have big uncertainties; therefore, it is not possible to accurately measure the spatial distance between two galaxies with photometric redshifts. 

The effect of photometric redshifts on the identifying mergers is studied at lower redshifts \citep[e.g. ][]{2015A&A...576A..53L}. In case of redshifts beyond 6 the main problem is likely to be not the uncertainty of photometric redshifts (if the probability density function of the redshift is known, one can marginalize over it), but the so-called catastrophic outliers, i.e. when a low-redshift dusty galaxy is misidentified with a high-redshift galaxy (see further discussion in \S\ref{sec:res:contamination}).

The area of the deepest part of the XDF\footnote{The corners of the XDF field are taken as [(03:32:45; -27$^{\circ}$ 46$'$ $44.95\arcsec$), (03:32:38; -27$^{\circ}$ 45$'$ $34.68\arcsec$), (03:32:32; -27$^{\circ}$ 47$'$ $13.72\arcsec$), (03:32:39; -27$^{\circ}$ 48$'$ $25.95\arcsec$)] as cited at \url{https://archive.stsci.edu/prepds/xdf/} }  (which is slightly smaller HUDF09) 
is $\sim1.9\arcmin\times2.1\arcmin$ that roughly corresponds to a $4\times4\times650\;(h^{-1}\mathrm{Mpc})^3$ pencil beam at redshifts from 6 to 9. The total number of detected sources in the field at all redshifts is $\sim7000$ \citep{Illingworth2013TheEver}; among them are 50 objects with redshifts $z\sim7$, and 22 with $z\sim8$ \citepalias{Bouwens2015UVFIELDS}. As we show in \S\ref{sec:results}, the former sample agrees well with the simulations and does not include any groups (defined by $1\arcsec$ threshold), while the latter sample does include groups and therefore it is of greater interest.

Among 22 galaxies at $z\sim8$, there are 3 groups (with 6 galaxies in total), where the distance between galaxies is less than $\sim1\arcsec$ (or $\sim30h^{-1}\mathrm{kpc}$). This separation criteria is quite arbitrary, and in this particular case it is motivated by the fact that there are no groups with separation $1-2\arcsec$ in this sample (in \S\ref{sec:results:2} and Figure \ref{fig:fraction_in_groups} we vary the separation threshold). In the Figure \ref{fig:cutout} these groups in different filters are presented. Also, there are a few cases where a galaxy with $z>5$ is in a close proximity, that can potentially be a clue for a misidentified redshift.

Even though we use the term ``merger'' throughout the paper, its definition is quite ambiguous. When two blobs are observed within $1\arcsec$ (projection corresponds to roughly $\sim30h^{-1}\mathrm{kpc}$) there is no guarantee that they are merging. Alternatively, they can be in a pre-merging state. Therefore, by ``mergers'' we mean the objects will occasionally merge.

Overall, it is safe to say that even the existing imaging data show the possibility of observing mergers at high redshifts. However, it is hard to quantitatively estimate whether we underestimate or overestimate the total fraction of mergers. The foregrounds definitely decrease the number of observed high-redshift galaxies, and possible catastrophic photometric redshift errors can decrease as well as increase the total number of high-redshift galaxies and the merger fraction.

\subsection{Example of a group}
\label{sec:example}

We select one of the groups and study its photometric redshifts in more details in order to be sure that it indeed can be a merger. The system contains three objects, that we mark as A, B and C. To the best of our knowledge, none of these objects has a spectroscopic redshift. We adopt photometric catalog by 3D HST \citep{Momcheva2016THEGALAXIES} and rerun EAZY code to extract the probability density functions. We use the settings and the templates identical to those used in 3D HST except two things: (1) we increase the hard limit on the maximum redshift from 6 to 12; (2) we do not apply any prior on the luminosity function. In the Figure \ref{fig:sed_pz} the cutout of the system, photometry and the redshift probability density functions, $P(z)$, are presented.

The photometric redshifts for these galaxies in \citetalias{Bouwens2015UVFIELDS} catalog, $z_\mathrm{Bouwens}$, and our redshifts of best fit (smallest $\chi^2$), $z_a$, and weighted \textit{peak} redshift\footnote{See \citet{Brammer2008EAZY:Code} for details regarding $z_a$ and $z_{peak}$.}, $z_\mathrm{peak}$, are shown in Table \ref{tab:phot}. The values differ due to different photometry, templates, etc; however, the object A is identified as a lower redshift galaxy compared to B and C. The width of probability density functions of photometric redshifts (see bottom left panel in the Figure \ref{fig:sed_pz}) shows that in fact all three galaxies can be at the same redshift. The probability of a simple projection coincidence for two objects, B and C, being nearby is $\sim5\%$ (see \S\ref{sec:res:coins}). 

Alternative interpretation of the group is a strong lens. In this particular case, A is unlikely to be an image of the same object as B and C since it is distinctly visible in $F850LP$ and $F775$ filters, while B and C are not. Therefore, we explore a possibility of B and C being the images of the same object. The Figure \ref{fig:lens} shows what point mass at which redshift is necessary to achieve observed separation of $\sim1\arcsec$, and whether such a lens (in case if it is an elliptical galaxy) could be not detectable by the HST. The estimation of the sensitivity curve is similar to the one made in \citet{Tsai2015THEWISE}. The conclusion is that it is possible to ``hide'' the lens.

This system illustrates the uncertainty associated with identifying the nature of such groups. Here all three possibilities -- actual mergers, projection, strong lens -- can take place; however, as we will show in \S\ref{sec:results}, the mergers are a slightly more probable interpretation of observing multiple groups.

\begin{table}[]
\centering
\begin{tabular}{l|lll}
              & A    & B    & C    \\ \hline
$z_\mathrm{Bowens}$ & 5.82 & 7.49 & 7.49 \\
$z_{a}$      & 6.17 & 7.41 & 8.02 \\
$z_\mathrm{peak}$   & 6.17 & 6.55 & 8.01 \\ 
\end{tabular}

\caption{Photometric redshifts as listed in \citepalias{Bouwens2015UVFIELDS} catalog, and our estimates made with running EAZY \citep{Brammer2008EAZY:Code} on the 3D HST photometric catalog \citep{Momcheva2016THEGALAXIES}.}
\label{tab:phot}

\end{table}

\section{Numerical Methods}
\label{sec:methods}
\subsection{Numerical Simulations}
\label{subsec:numsim}

In order to properly model the number of mergers and their separation we use a numerical simulation with all the necessary physics -- hydrodynamics and star formation -- included, and where the times of merging are fully modeled. 
Alternatively, one can perform the same study with a dark matter only simulation or even just with a mock halo catalog and a halo occupation distribution (HOD) model. However, here the dynamic timescales that correspond to the merging of galaxies, not only dark matter halos, might be important. While these timescales can be considered analytically, e.g. \citet{Simha2016ModellingDestruction} or \citet{Hong2016TheTimescales}, still, such methods are approximate.

We adopt the suite of simulations of the cosmic reionization \citep{gnedin_cosmic_2014-1,gnedin_cosmic_2014}, in which the spatial resolution of the adaptive grid reaches 100 proper pc. We use a $40h^{-1}$Mpc box, for which we have only a few snapshots, and a series of ten $10h^{-1}$Mpc simulations (with the initial conditions drawn randomly with a random DC mode \citep{Gnedin2011ImplementingVariables}). Smaller boxes allow us to have better resolution in time, while the larger box is used to confirm our results. The $40h^{-1}$Mpc box is relatively small comparing to some other modern simulations, but still has a greater statistical power than the volume of the XDF.

\subsection{Mimicking the observations}
\label{subsec:mimobs}

\begin{figure}[!t]
\includegraphics[width=0.99\columnwidth]{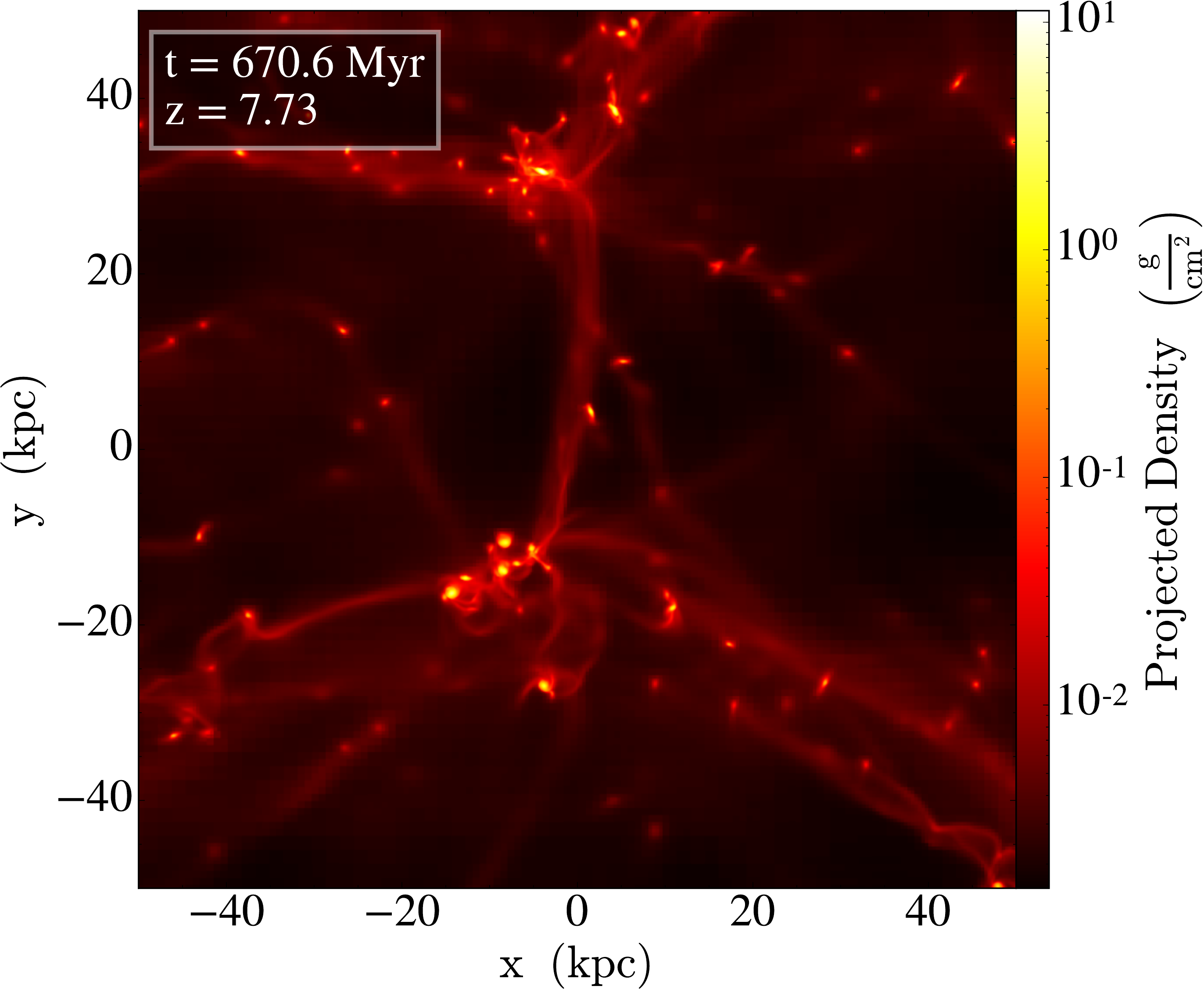} 
\caption{\label{fig:gas_projection}Projection of gas density at redshift 7.7 in a 100 kpc cube. The region corresponds to the proximity of the brightest galaxy at this redshift in the $80h^{-1}$Mpc box.}
\end{figure}

\begin{figure}[!t]
%Panel 1: gas projection \\
\includegraphics[width=0.99\columnwidth]{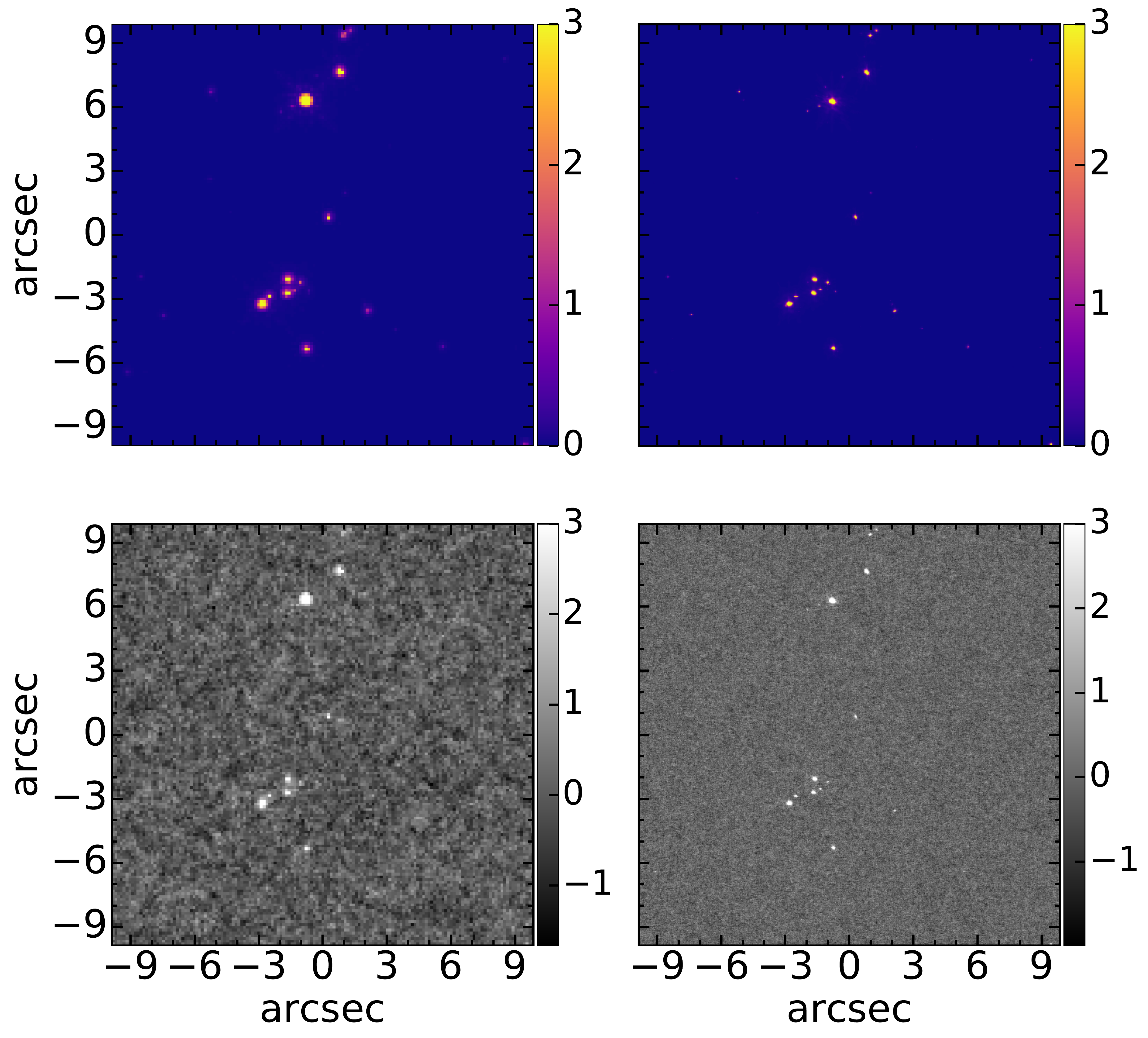} 
\hspace*{0.53cm} \includegraphics[width=0.42\columnwidth]{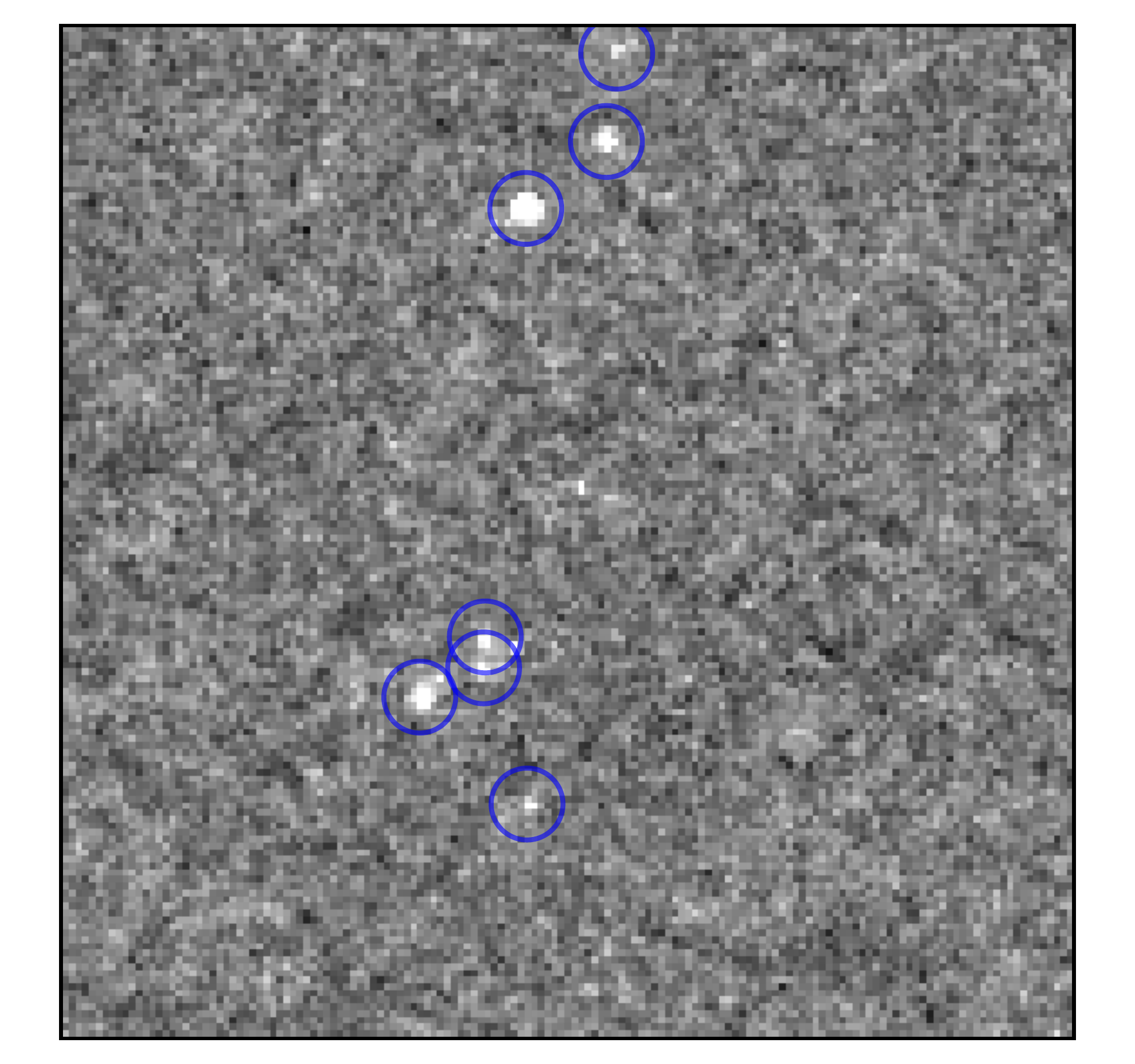} \hspace{0.15cm}
\includegraphics[width=0.42\columnwidth]{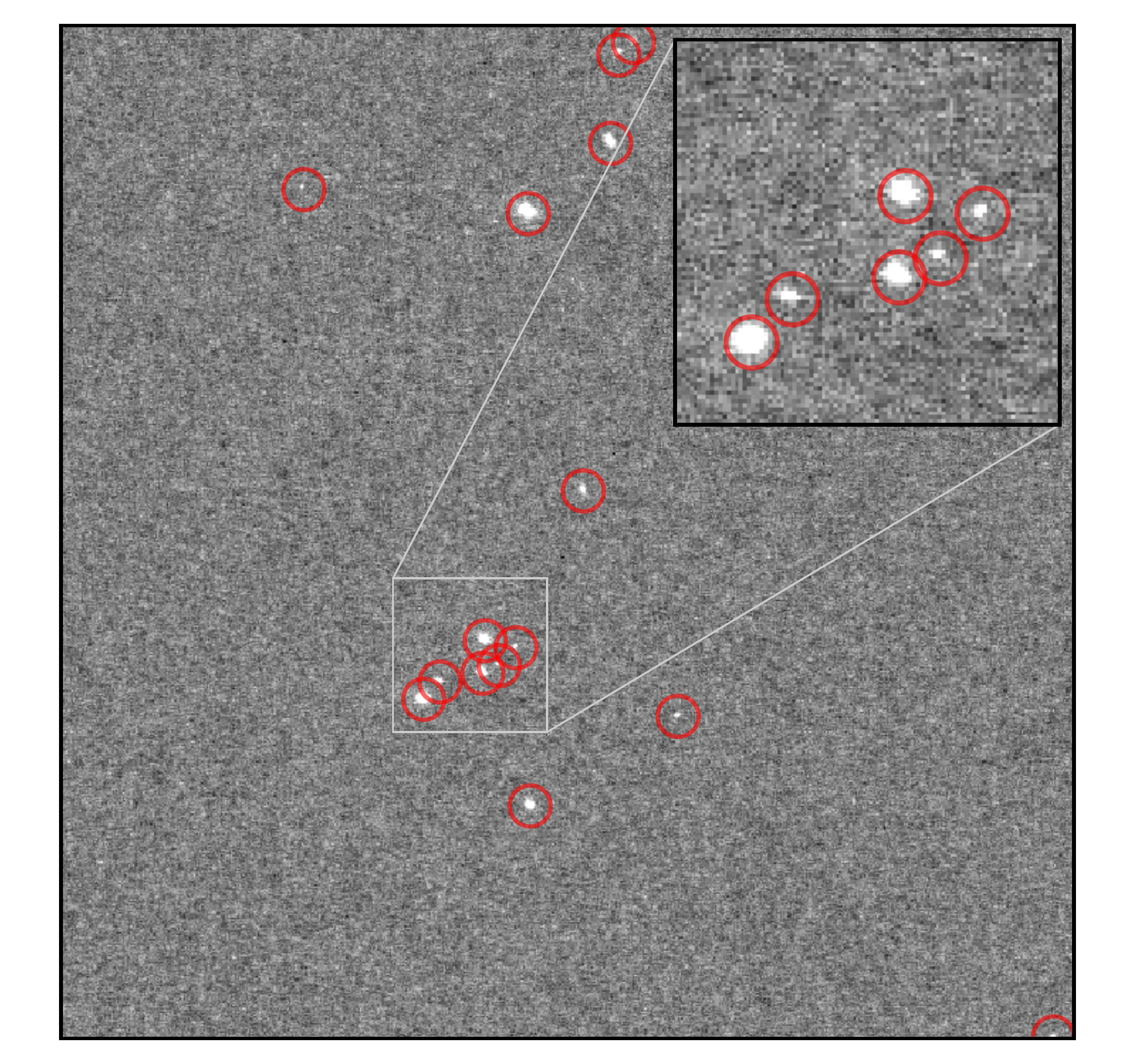} \\
 
\caption{\label{fig:image_generation_step1}  
Steps of generating mock observation for the F160W filter WFC3 HST (left panels) and the F150W filter  NIRCam JWST (right panels). \textit{First row:} Same projection as in Figure \ref{fig:gas_projection} for star particles, convoluted with corresponding point spread functions in units of sigma (relative to the noise). \textit{Second row:} The final mock observation with corresponding noise added (see \S\ref{subsec:mimobs} for details). \textit{Third row:} identified sources with ``photutils'' package using combined image in 3 filters ($F125W$, $F140W$ and $F160W$ filters for the HST and $F115W$, $F150W$ and $F200W$ filters for the JWST).}
\end{figure}

Since the simulations incorporate star formation, we can calculate the fluxes directly. In order to properly evaluate the flux for each filter, we first calculate the spectrum for each star particle using Flexible Spectral Population Synthesis (FSPS) code \citep{conroy_propagation_2009}, and then apply appropriate redshift and the IGM absorption corrections \citep[we adopt an updated \citet{Madau1995RadiativeGalaxies} model by][]{inoue_updated_2014}. The dust extinction of the dim high-redshift galaxies is expected to be low at $z\gtrsim7$ \citep[e.g. ][]{Bouwens2014UV-CONTINUUMFIELDS}. The bright galaxies are affected more; however, it does not suppress their detectability. Therefore, in this study we assume the dust effect to be negligible.

Initially we produce a projection with a resolution higher than the resolution of the detectors by factor of a few, then we apply the point spread function corresponding to a given filter and instrument.

In order to mimic the noise of the HST, we manually extract the empty regions (without any objects detected) from the legacy data of the deepest of the deep fields -- XDF \citep{Illingworth2013TheEver}. Then, we evaluate the noise pattern (the amplitude and smoothness, that correspond to the point spread function) that we later use for generating noise. Alternatively, we tried to use the noise cutouts and got identical results. 

For the JWST we used the demo version of ``JWST Exposure Time Calculator'' that allows to generate noise for a given filter, observation time, background level, readout pattern. For our estimates we used 30 hours observations per filter and a ``low'' background noise. We found that other parameters do not significantly affect our results. A detailed study of different parameters and strategies is beyond the scope of this paper.

In the final step we downsample our projections and add them to the noise. We show the projection of the gas density in a 100 kpc box at $z=7.7$ in Figure \ref{fig:gas_projection}, and then the corresponding mock observation by HST and JWST in Figure \ref{fig:image_generation_step1}. For illustrative purposes, the region shown in the figures is selected to be very overdense and corresponds to the proximity of the brightest galaxy in $80h^{-1}\rm{Mpc}$ box at $z=8$. Since $80h^{-1}\rm{Mpc}$ box is almost 100 times larger than the effective XDF volume (in the range $7<z<10$), such a system is unlikely to be among observed.

\subsection{Processing the observations}
\label{subsec:proc}

In order to extract the flux from an image we adopt widely used tool ``SExtractor'' \citep{Bertin1996SExtractor:Extraction} and an alternative instrument ``photutils'' \citep[an associated package of ``astropy'', ][]{Bradley2016Astropy/photutils}, which provides similar functionality. These software allow one to find an object on a noisy image, determine its size and shape, and then derive its flux. We tried both methods and got similar results when used with similar parameters. We found our results to be more sensitive to the choice of parameters rather than the software. The results presented below are made with use of ``photutils''\footnote{The ``photutils'' package provides the functionality of separating objects with multiple blobs into individual, which we did not use in this study.}.

The objects were detected in the combination of F125W, F140W and F160W filters for the HST, following the common practice (e.g. 3D HST \citet{Skelton20143D-HSTMASSES}), and in the F115W, F150W and F200W combination for the JWST. The result of the source detection is shown in the Figure \ref{fig:image_generation_step1}. It is immediately apparent that the JWST is superior and in 30 hours of observation is capable to improve over the HST's deepest image. \added{The number count mostly increases because of the depth, and the smaller point spread function accounts only for $\sim1\%$ of additional detected objects.}

Finally, we take into account that drop-out samples $z\sim7$ and $8$ have some distribution in redshift space \citep[see Figure 4 in ][]{Bouwens2011ULTRAVIOLETOBSERVATIONS}. To do so, we converge this distribution with the redshift dependent results we got from the numerical simulation.

\begin{figure}
\begin{center}

  \begin{overpic}[width=0.99\columnwidth]{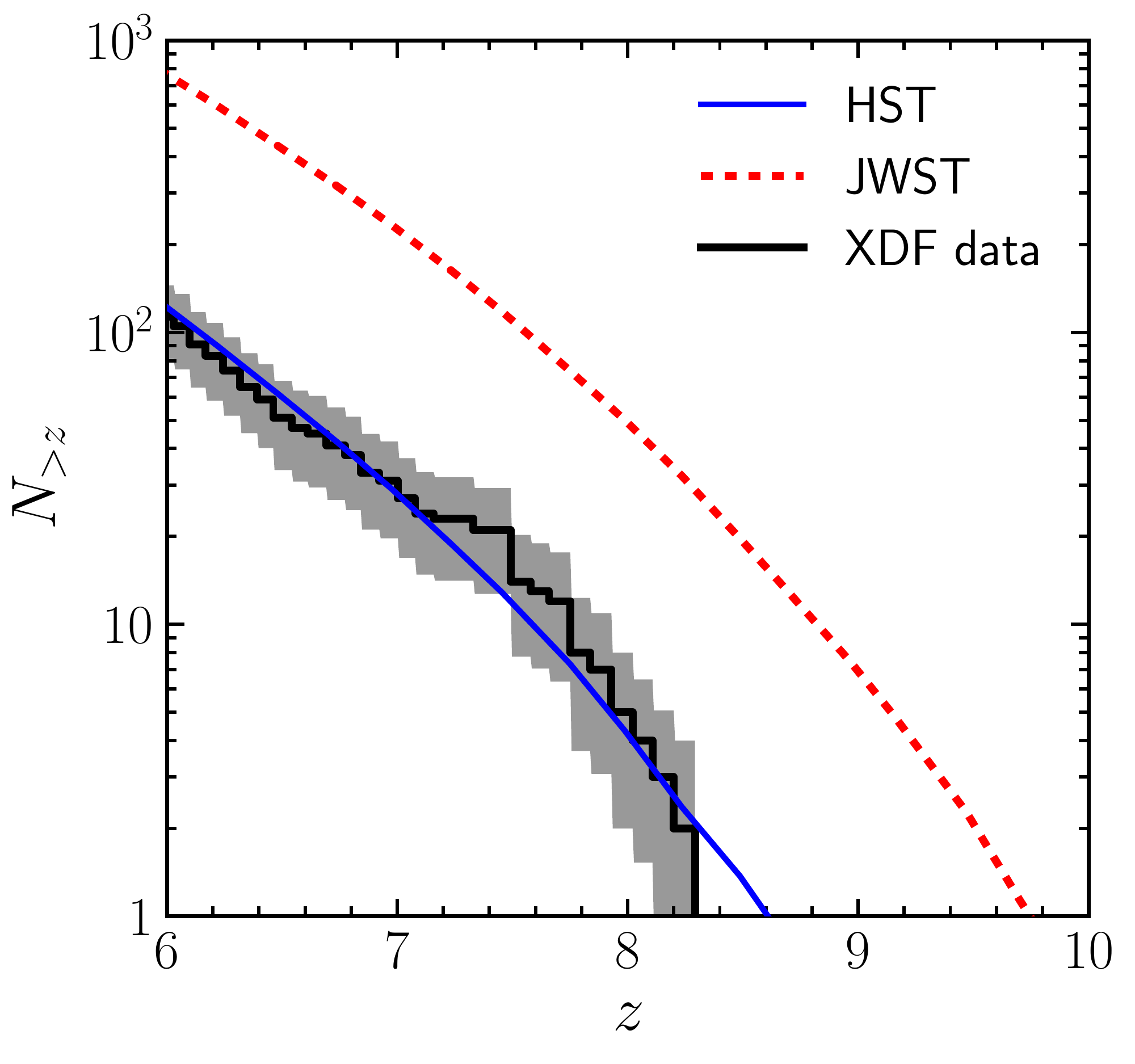}
  
  \end{overpic} 
\caption{\label{fig:sources} Number of objects in the XDF field at redshifts above $z$ (black curve) with scatter corresponding to the Poisson noise and cosmic variance (shaded region). For this figure only we adopted the photometric redshifts derived with template fitting code EAZY from \cite{Bouwens2015UVFIELDS}, while in the rest of the paper we use drop-out samples from the same paper. Blue curve is the result of the forward modeling mimicking the depth of the XDF, therefore it matches the data. Red curve is the forecast of the number of objects the JWST will be capable to observe in the same field (assuming 30 hours integration per each of these filters: $F115W$, $F150W$ and $F200W$).
}
\end{center}
\end{figure}

In Figure \ref{fig:sources} the total number of observed sources is plotted. The simulations do match the observations as expected since the adopted simulation has been already shown to match the observed luminosity functions of galaxies combined from multiple deep fields \citep{gnedin_cosmic_2014}. This agreement confirms the legitimacy of the approach.

The foreground galaxies obscure some parts of the field. Therefore, the XDF data that we present here underestimates the actual number of galaxies. Also, notice that due to the photometric redshift uncertainties there should be scatter along x-axis, which is not included here.

Even though we tried to follow the source extraction approaches as used in the 3D HST survey, we still had some flexibility in the parameters. This flexibility can cause $10-20\%$ change in the number of the detected sources. We chose parameters to reproduce the observed number counts identically, therefore the match with the data in the Figure \ref{fig:sources} is almost perfect.

\section{Results}
\label{sec:results}

\added{We have shown that there are groups of galaxies in drop-out $z\sim8$ sample in \citet{Bouwens2015UVFIELDS} catalog, while there are none of such groups in $z\sim7$ sample. We summarize all possible explanations in the subsections below and the corresponding probabilities are listed in the Table \ref{tab:sum}. Overall, we show that while all of three mechanisms are unlikely to individually account for all of three pairs, a combination of them can, and some of the observed pairs are likely to be real mergers.}

\begin{table}[]
\centering
\begin{tabular}{l|lll}
Physical effect                                                               & 1 pair        & 2 pairs        & 3 pairs          \\ \hline
Projection coincidence (\ref{sec:res:coins})                                                   & 5\%           & 0.1\%          & 0.001\%          \\
Strong lensing (\S\ref{sec:res:lens})  & \textless20\% & \textless0.2\% & \textless0.002\% \\
Mergers (\S\ref{sec:results:2})                                                                       & 15\%          & 4\%            & 0.2\%            \\
Catalog contamination (\S\ref{sec:res:contamination})                                                         & \multicolumn{3}{c}{unknown}                      
\end{tabular}
\caption{Probability of different mechanisms to account for different number of pairs in the XDF at $z\sim8$. Groups of three objects are neglected, but for each mechanism it is a much lower probability.}
\label{tab:sum}
\end{table}

\added{
\subsection{Projection coincidences}
\label{sec:res:coins}
In order to estimate the projection coincidences we run a Monte Carlo simulation by randomly populating a $2.1\arcmin\times1.9\arcmin$ field with 22 points and calculating the number of points with a neighbor within $1\arcsec$. The results for $0, 2, 3, 4, 5$ and  $6$ objects with pairs is $95\%, 4.8\%, 0.012\%, 0.1\%, 0.0004\%$ and $0.001\%$. Therefore, all three pairs are highly unlikely to be coincidences. Projection coincidences can be reduced with better photometry and completely eliminated with deeper spectroscopic surveys. 
}

\subsection{Lensing}
\label{sec:res:lens}

In \S\ref{sec:example} we showed that a ``hidden'' lens can produce the pair of images of a $z\sim8$ galaxy. Here we discuss what is the probability of such an event.

The lensing effect has been studied in \citet{Wyithe2011ALensing}, and its effect on the luminosity function in \citet{Mason2015CORRECTINGBIAS,Fialkov2015DISTORTIONLENSING,Barone-Nugent2015TheFields}. The general conclusion is that it is significant and should be taken into account for the luminosity function derivation. However, the multiple imaging, which is of interest in this paper, is less definite since it partially relies on the faint-end slope of the luminosity function which is not well measured. 

In \citet{Wyithe2011ALensing} the fraction of observed multiply imaged systems at $z\gtrsim7$ is estimated to be order of $\lesssim1\%$; therefore we can roughly estimate the probability of 3 objects to be lensed as $20\times1\%^3=0.002\%$, i.e. same order of magnitude as the projection coincidences.

\subsection{Merger statistics}
\label{sec:results:2}

\begin{figure}
\includegraphics[width=0.99\columnwidth]{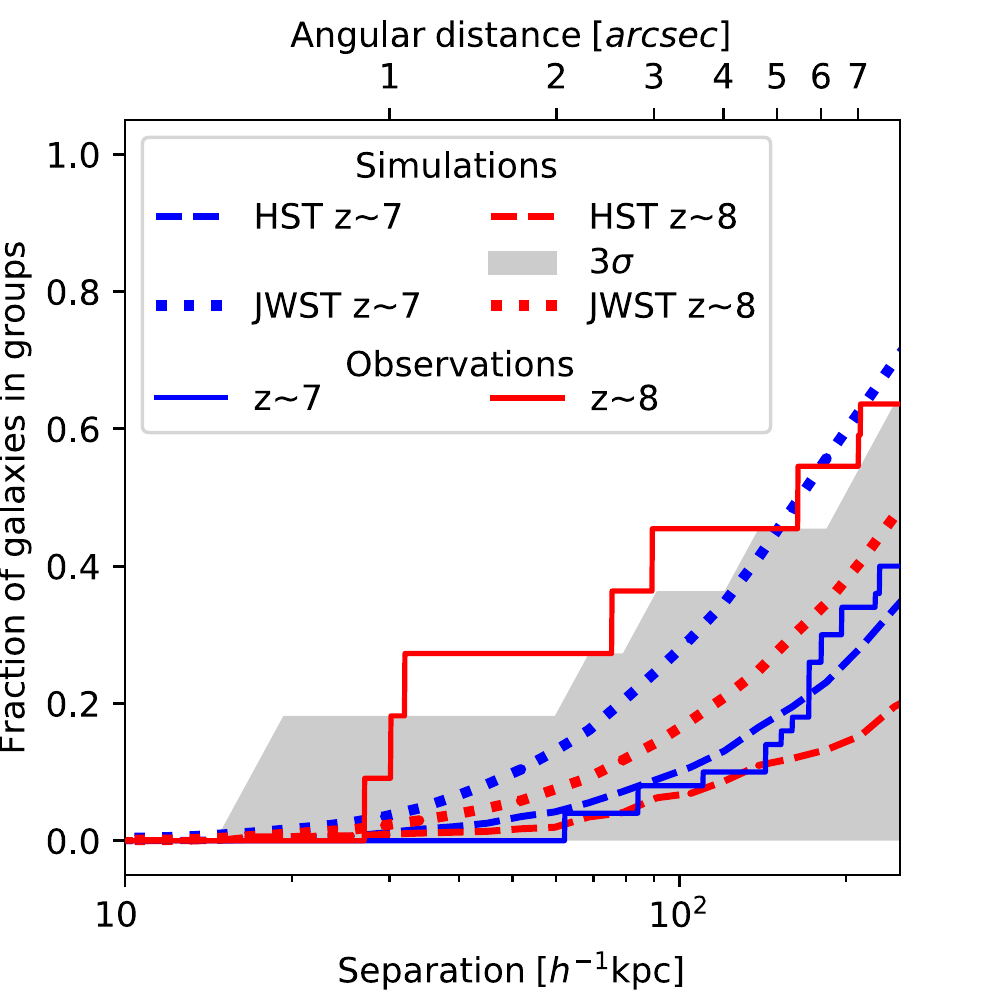}
\caption{\label{fig:fraction_in_groups}
Cumulative fraction of galaxies that have a companion within given angular distance (distance in $h^{-1}$kpc given for $z=7.5$) for $z\sim7$ and $z\sim8$ drop-out samples along with the results from simulations. The error region ($99.7\%$ region) is shown only for the simulation result at $z\sim8$ for the HST, and it is estimated with Monte Carlo approach. The error is underestimated due to the neglected cosmic variance (see Appendix). The projection of the JWST measurements as calculated assuming 30 hours integration per each of these filters: $F115W$, $F150W$ and $F200W$.
}
\end{figure}

In Figure \ref{fig:fraction_in_groups} the fractions of objects with a companion within given radius for $z\sim7$ and $z\sim8$ are presented. The $z\sim7$ sample is consistent with the simulations. However, three groups of galaxies in $z\sim8$ sample cause a noticeable divergence. 

The error region for the simulated $z\sim8$ sample in Figure \ref{fig:fraction_in_groups} is calculated using Monte Carlo approach by generating many mock observations. The discrepancy barely reaches the significance of $3\sigma$. Since some of these pairs can be projection coincidences, we can not report this as a significant discrepancy. In order to definitively say whether it is an anomaly, one has to improve the statistics. We believe that it is even possible to do with existing data; our preliminary results show that similar groups of galaxies can be found in other deep fields. However, in this study we limited our exploration to the XDF data, because the inconsistency in the number and the depth of the filters of other deep fields complicates the analysis.

In \citet{Park2017Dark-agesGalaxies} similar comparison between numerical simulation and observations are made. In contrast to this paper where we use the merger fraction, the authors adopt the angular correlation function. Also, in that study other deep fields are included into consideration in addition to the XDF data, and redshifts are considered starting from 6 and above. The authors find a good agreement with simulation. 

In Figure \ref{fig:fraction_in_groups} the merger fraction for the JWST is also plotted. As expected, the fraction of observed mergers is higher compared to the HST. The main caveat is likely to be the confusion limit. The total number of sources will greatly increase (almost by factor of 10), including foreground dim galaxies, while the point spread function of the JWST is not dramatically superior than HST's. Therefore, the field will be crowded with galaxies and our criteria for mergers -- within certain angular distance and in the same photometric redshift bin -- may not be sufficient, and the probabilistic methods would be required.

\subsection{Catalog contamination}
\label{sec:res:contamination}
\begin{figure*}
\centering
\includegraphics[height=120pt]{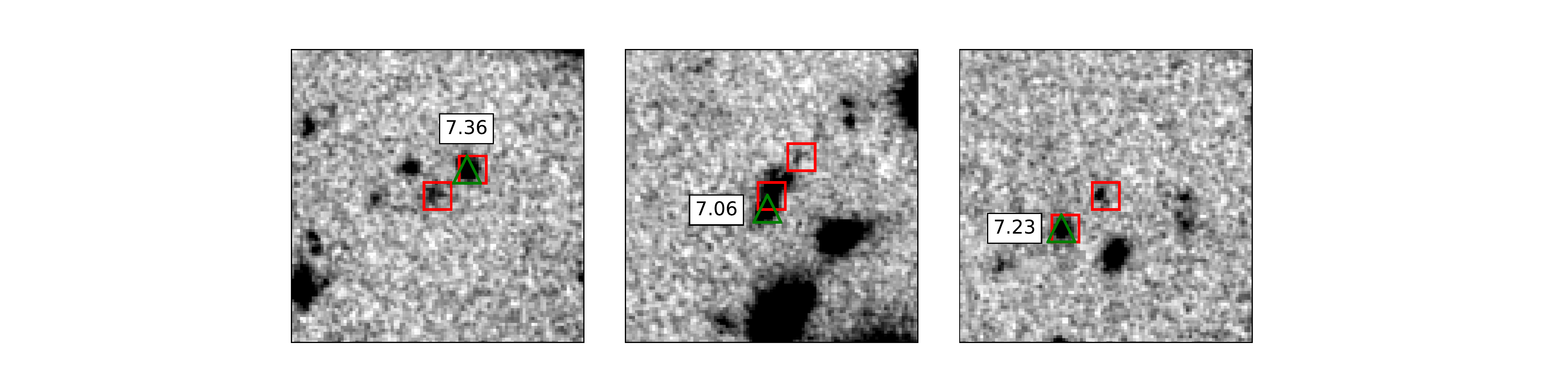}
\caption{\label{fig:mclure}Cutouts in $F160W$ filter identical to those in Figure \ref{fig:cutout}. Red squares correspond to the galaxies in $z\sim 8$ sample from \citet{Bouwens2015UVFIELDS}, green triangles and labels correspond to the galaxies and their photometric redshifts in \citet{Mclure2011AMasses} catalog.}
\end{figure*}

\added{Finally, there is a possibility that the observed groups in $z\sim8$ sample are caused by the contaminations from lower redshifts.  The absence of a sufficiently large spectroscopic sample at the redshifts of our interest makes it impossible to perform a similar quantitative analysis. Nevertheless, it is often assumed that the contamination is low, following the arguments presented in  \citet{Bouwens2011ULTRAVIOLETOBSERVATIONS} (see \S4.2 therein).

The catalog by \cite{Mclure2011AMasses} has a much lower number of high-redshift galaxies (see discussion and detailed comparison with \citet{Bouwens2011ULTRAVIOLETOBSERVATIONS} therein). There are 8 galaxies in the region we consider with $z>7$, and all of them are present in $z\sim 7$ or $8$ samples of \citetalias{Bouwens2015UVFIELDS}. Only 6 out of 22 objects of $z \sim 8$ have a counterpart in \citet{Mclure2011AMasses} catalog; however, there is one of these 6 galaxies in each of three groups we studied (see Figure \ref{fig:mclure}). Therefore, the tension between the number of groups in \citetalias{Bouwens2015UVFIELDS} catalog and our theoretical predictions might be interpreted as an argument in favor of a more conservative catalog like \citet{Mclure2011AMasses}.}

\section{Discussion}
\label{sec:discussion}

\deleted{We have shown that the mergers at $z\gtrsim7$ are already can be observed in the photometric surveys, and with the upcoming JWST data, their fraction will dramatically increase. Moreover, while individual groups are almost indistinguishable from the multiply imaged systems by strong lensing and from the projection coincidences, their predicted fraction is an order of magnitude larger. }

\added{Under the assumption, that some fraction of the observed groups are real mergers, we discuss how they could be further studied and used.} Most of the applications listed below are already applied to the lower redshift data.

\added{Since the distribution of mergers is more biased than isolated galaxies (see Appendix, and \citet{Furlanetto2005PairBias} for an analytical model), the merger fraction can be used as the indicator of the local overdensity and, consequently, it may allow to control the cosmic variance effect on the luminosity function. For instance, this study shows that the XDF field is likely to correspond to an overdense region at $z\sim8$.}

As was mentioned in the introduction, there are a couple of LAE mergers detected at $z>6$. Since Ly$\alpha$ is one of a few spectroscopic emission lines that can be observed at high redshifts, it is often the only way to measure spectroscopic redshift for reionization galaxies. Also, the studies of Ly$\alpha$ emitters and their morphology at lower redshifts ($z\approx 4$) confirm the models which suggest that merging and close encounters can activate Ly$\alpha$ emission \citep{kobayashi_morphological_2016}. Therefore, mergers can be potential candidates for Ly$\alpha$ follow up observations.

Beside the connection between mergers and LAEs, the statistics of mergers at high redshifts may shed light on other questions. Among them is the accordance between dark mater halos and galaxies. While the luminosity function of galaxies allows to perform abundance matching with dark matter halos, the merging statistics contains additional information that allows to perform more complicated analysis and study assembly bias and subhalo abundance matching (SHAM). Recently, a comprehensive study of Sloan galaxies at lower redshifts, where statistics is orders of magnitudes better, has been done in \citet{Guo2016ModellingMatching}.

It is known that at lower redshifts the major mergers enhance star formation, while at intermediate redshifts ($z\sim1-3$) the effect seems to be less prominent \cite{Fensch2016High-redshiftFormation}. Depending on how strong the effect of the enhanced star formation at the epoch of reionization, the stacking methods like \citet{Geil2017DarkDawn} might be able to detect larger ionization bubbles with 21 cm signal around such systems.

The galaxy mergers may also improve photometric redshifts. The galaxies identified within the same merging group allow one to consider them to be at the same redshift, and therefore improve their photometric redshift by increasing precision (combined signal has larger signal-to-noise ratio) and accuracy (avoiding catastrophic errors using prior of the abundance of mergers at different redshifts). The spectroscopic redshifts matched with some members of the groups can further improve the measurements \citep{Ho2008CorrelationImplications}. Significant amount of work has been already done to incorporate the spatial correlation of galaxies into determination of photometric redshifts at lower redshifts \citep[e.g.][]{Newman2008CalibratingCross-Correlations, McQuinn2013OnDistributions,Menard2013Clustering-basedData}, and these techniques can be applied to high-redshift galaxies.

Moreover, the uncertain but identical redshifts of galaxies within a merging group will allow one to directly compare their photometric SEDs. Such a comparison will provide information regarding the UV slopes and the dust abundance as function of galaxy mass.

\software{SExtractor \citep{Bertin1996SExtractor:Extraction}, photutils \citep{Bradley2016Astropy/photutils}, yt \citep{Turk2011}, EAZY \citep{Brammer2008EAZY:Code}.}

\acknowledgments
AAK is supported by the Friends of the Institute for Advanced Study. This work is based on observations taken by the 3D-HST Treasury Program (GO 12177 and 12328) with the NASA/ESA HST, which is operated by the Association of Universities for Research in Astronomy, Inc., under NASA contract NAS5-26555.

\appendix
\twocolumngrid
\added{\section{Rescaling the simulation}}
\label{app:scaling}

In the main body of the paper we assumed that the observed volume is a typical region of the universe. However, it might not be the case, for instance, if by chance the XDF region happen to correspond to a few sigma overdensity at redshift $z\sim8$. In this appendix we explore how this possibility alters our chances to observe the pairs of galaxies.

In order to mimic overdense or underdense regions we rescale the simulation by linearly shifting the time. Since we are interested in relatively short time interval that corresponds to the $z\sim8$ drop-out sample, rescaling in redshift or scale parameter space gives the same result. After rescaling the simulation does not longer match the observed luminosity functions on lower redshifts; however, it is not relevant for this particular numerical experiment.

For each time shift of the simulation we use Monte Carlo method to generate thousands of mock observations. Firstly, we randomly stack our simulation box in order to fill the redshift range corresponding to $z\sim8$. Secondly, we generate pencil beams that match the XDF volume and $z\sim8$ selection function \citep{Bouwens2011ULTRAVIOLETOBSERVATIONS}. In result we have multiple realizations some of which contain more or less than 22 objects and some number of pairs. We select only those realizations that have exactly 22 objects and record the probability of this event in terms of $\sigma$'s of normal distribution.. For convenience we use this $\sigma$ to parametrize the rescaling of the simulation. With this notation the original simulation without rescaling corresponds to roughly zero $\sigma$, because it was tuned to match the observation. In case if we delay it, the typical number of objects drops and an observation with 22 galaxies corresponds to a few $\sigma$. 

\begin{figure}
\includegraphics[width=0.99\columnwidth]{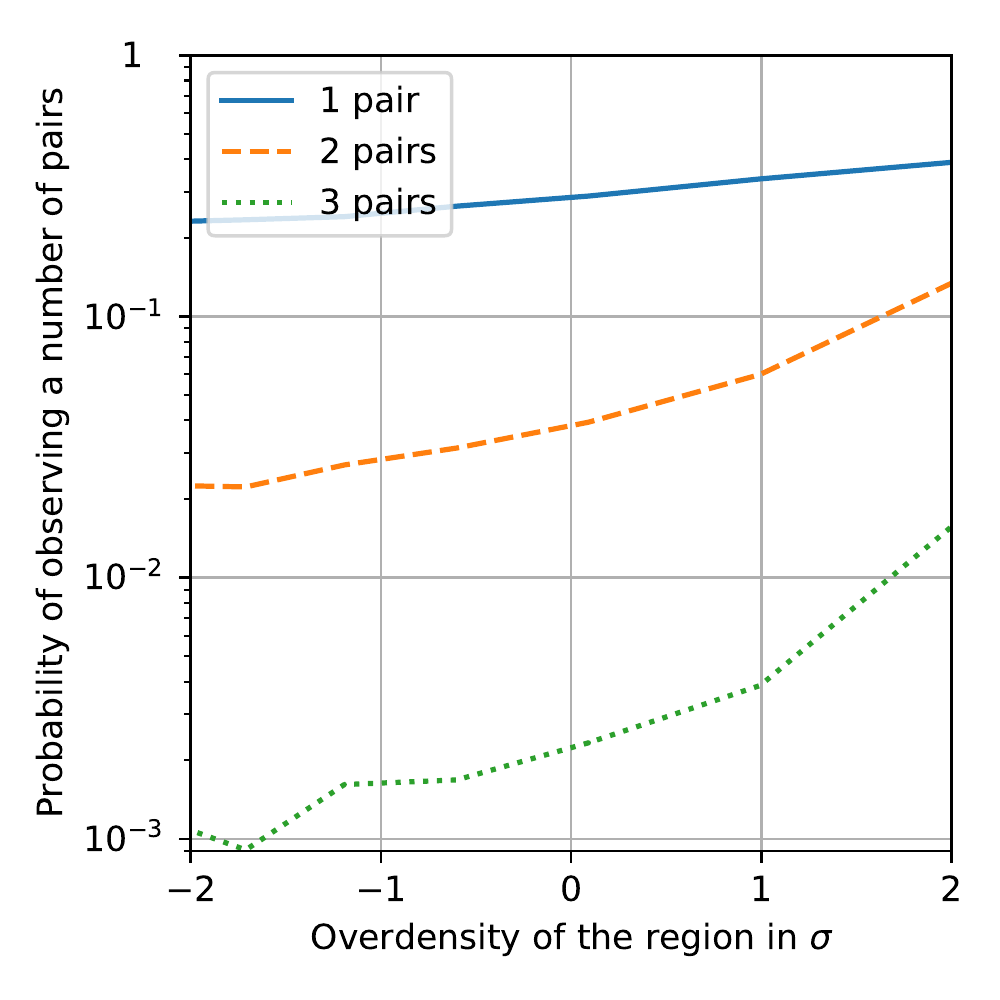}
\caption{\label{fig:sigmamap}Probability of observing certain number of pairs in a mock $z\sim8$ drop-out survey after we rescaled a simulation. The figure shows that we See the Appendix for the details regarding rescaling the simulation.
}
\end{figure}

Also, we keep record of how many pairs (in this case defined by $1\arcsec$ separation criteria) were present in each realization with 22 objects. In Figure \ref{fig:sigmamap} we show the fraction of realizations that have certain number of pairs versus rescaling parameter $\sigma$. It can be seen, that when we delay our simulation and, therefore, consider an overdense region, we are more likely to see multiple pairs, and observing 3 pairs is not longer an unlikely event. The numbers presented in Table \ref{tab:sum} correspond to the $\sigma$ equals zero in the Figure \ref{fig:sigmamap}.

The trends in the Figure \ref{fig:sigmamap} are the result of the fact that mergers are more biased than `field' galaxies. See \citet{Furlanetto2005PairBias} for an analytical model that describes this effect.

\bibliographystyle{yahapj}
\bibliography{Zotero.bib,Mendeley.bib,refs.bib}

\end{document}